\documentclass[sigconf]{acmart}
\usepackage{multirow}
\usepackage{array}

\newcommand{\rev}[1]{\textcolor{black}{\begingroup#1\endgroup}}

\AtBeginDocument{%
  }

\copyrightyear{2026}
\acmYear{2026}
\setcopyright{cc}
\setcctype{by}
\acmConference[IUI '26]{31st International Conference on Intelligent User Interfaces}{March 23--26, 2026}{Paphos, Cyprus}
\acmBooktitle{31st International Conference on Intelligent User Interfaces (IUI '26), March 23--26, 2026, Paphos, Cyprus}
\acmPrice{}
\acmDOI{10.1145/3742413.3789154}
\acmISBN{979-8-4007-1984-4/2026/03}

\usepackage[nolist]{acronym}
\begin{acronym}
    \acro{HCI}{Human-Computer Interaction}
    \acro{LLM}{Large Language Model}
    \acro{CS}{Computer Science}
    \acro{RAG}{Retrieval-Augmented Generation}
    \acro{AI}{Artificial Intelligent}
    \acro{UI}{User Interface}
    \acro{MMR}{Maximal Marginal Relevance}
    \acro{MMRAG}{Multimodal Retrieval-Augmented Generation}
    \acro{MLLM}{Multimodal Large Language Model}
\end{acronym}
\usepackage{dirtytalk}
\usepackage[draft]{fixme}
\usepackage{soul}

\begin{document}

\author{Junling Wang}
\authornote{These authors contributed equally to this work.}
\orcid{0000-0002-4526-2907}
\affiliation{%
  \institution{ETH Zurich}
  \city{Zurich}
  \country{Switzerland}
}
\email{junling.wang@ai.ethz.ch}

\author{Hongyi Lan}
\authornotemark[1]
\orcid{0009-0001-6550-474X}
\affiliation{%
  \institution{ETH Zurich}
  \city{Zurich}
  \country{Switzerland}
}
\email{honlan@student.ethz.ch}

\author{Xiaotian Su}
\orcid{0009-0004-0548-1576}
\affiliation{%
  \institution{ETH Zurich}
  \city{Zurich}
  \country{Switzerland}
}
\email{xiaotian.su@inf.ethz.ch}

\author{Mustafa Doga Dogan}
\orcid{0000-0001-8787-8681}
\affiliation{%
  \institution{Adobe Research}
  \city{Basel}
  \country{Switzerland}
}
\email{doga@adobe.com}

\author{April Yi Wang}
\orcid{0000-0001-8724-4662}
\affiliation{%
  \institution{ETH Zurich}
  \city{Zurich}
  \country{Switzerland}
}
\email{april.wang@inf.ethz.ch}

\renewcommand{\shortauthors}{Junling Wang et al.}
\newcommand{\AW}[1]{\textcolor{blue}{\textbf{*April*}: #1}}
\newcommand{\JW}[1]{\textcolor{purple}{\textbf{*Junling*}: #1}}
\newcommand{\XS}[1]{\textcolor{red}{\textbf{*Xiaotian*}: #1}}
\newcommand{\MS}[1]{\textcolor{purple}{\textbf{*Mrinmaya*}: #1}}

\newcommand{\inlinequote}[1]{``\textit{#1}''}

\begin{abstract}


Designing user interfaces (UIs) is a critical step when launching products, building portfolios, or personalizing projects, yet end users without design expertise often struggle to articulate their intent and to trust design choices. Existing example-based tools either promote broad exploration, which can cause overwhelm and design drift, or require adapting a single example, risking design fixation. We present \sys{}, an interactive system that supports mobile UI design through an example-driven design workflow. Powered by a multimodal retrieval-augmented generation (MMRAG) model, \sys{} enables iterative search, selection, and adaptation of examples at both the global (whole interface) and local (component) level. To foster trust, it presents source transparency cues such as ratings, download counts, and developer information. In an empirical study with 24 end users, \sys{} significantly improved participants' ability to achieve their design goals, facilitated effective iteration, and encouraged exploration of alternative designs. Participants also reported that source transparency cues enhanced their confidence in adapting examples. Our findings suggest new directions for AI-assisted, example-driven systems that empower end users to design with greater control, trust, and openness to exploration.


\end{abstract}

\newcommand{\sys}{\textsc{UI Remix}}
\newcommand{\condone}{RAG(source)}
\newcommand{\condtwo}{RAG(no source)}
\newcommand{\condthree}{GPT-4o}
\newcommand{\condfour}{Browsing}
\newcommand{\commu}{online learning communities}
\title[\sys{}: Supporting UI Design Through Interactive Example Retrieval and Remixing]{\sys{}: Supporting UI Design Through Interactive Example Retrieval and Remixing}



\begin{teaserfigure}
    \centering
    \includegraphics[width=\textwidth]{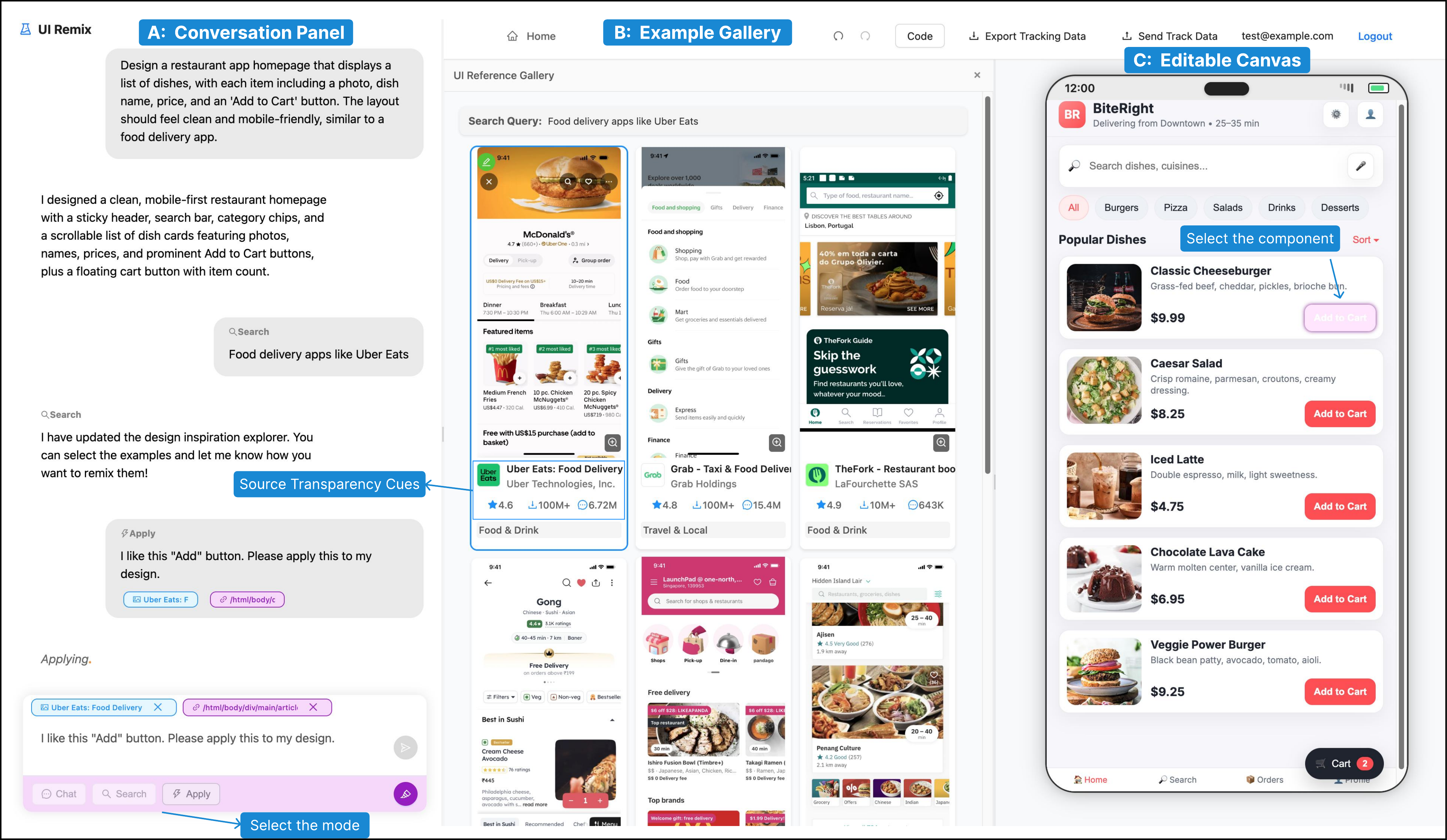}
    \vspace{-10pt}
    \caption{
    An overview of \sys{}, an example-driven assistant for mobile UI design featuring three main panels: (A) \textbf{Conversation Panel}: Users describe their design goals and interact with the system through three modes: Chat (generate and refine UI designs), Search (retrieve relevant UIs for inspiration), and Apply (adapt entire design through \textit{global remix} or specific components through \textit{local remix} from selected examples). 
    (B) \textbf{Example Gallery}: Displays retrieved real-world UI examples along with source transparency cues (e.g., ratings, download counts, developer information) to help users assess credibility. (C) \textbf{Editable Canvas}: Presents a live preview of the current design, supporting toggling between visual and code views.}
    \label{fig:fig1}
    \vspace{1em}
\end{teaserfigure}

\maketitle
\section{Introduction}

Designing \acp{UI} requires translating abstract ideas into concrete visual and interactive forms, a skill that traditionally demands both creative expertise and technical proficiency with design software.
However, an increasing number of \textit{end-user designers}, such as small business owners, students, hobbyists, and content creators, now create UIs using AI-powered tools~\cite{min2025,kim2022}.
For these users, interface design is not a specialized profession but an expressive means to communicate ideas, promote services, or prototype concepts quickly~\cite{wang2022end}.

While AI assistance has lowered the barrier to producing interfaces, it has not eliminated the difficulty of articulating what one wants to create. 
End-user designers can easily generate layouts, but they often struggle to specify visual intent beyond broad impressions \rev{~\cite{brickify2025, lyngsTellMeWhat2018, metzler2006beyond}}. 
For instance, a small business owner might know they want a ``modern'' or ``user friendly'' UI but find it difficult to translate these abstract impressions into specific design choices.
\rev{Recent LLM-based design tools show promise in translating loosely specified prompts into plausible initial designs~\cite{genuistudy2025}. However, even when an AI system produces a design that matches such prompts or generates a rationale explaining why the design is \say{modern} or \say{user-friendly}, end-user designers may still struggle to judge its quality.}
Lacking the design literacy to critically evaluate or justify aesthetic choices, they may hesitate to trust the system's output or commit to a particular direction\rev{~\cite{takaffoli2024,liuUnakiteScaffoldingDevelopers2019}}.

A more intuitive way to build confidence is through \textit{socially grounded evidence} rather than abstract AI rationales. 
When users can see where a design comes from --- who made it, how it has been received, and in what context --- it provides a familiar and trustworthy signal of quality, much like trusting a product recommended by a respected expert.

A promising direction to overcome these challenges is to leverage real-world UI examples. 
These examples can help users articulate their design intent by offering concrete references that make abstract preferences more tangible and comparable.
Presenting traceable examples to users could also enhance trust in design decisions by enabling users to evaluate these choices through transparent source information~\cite{ehsan2021,stuart2012social}.
Using existing UI examples to support UI design has gained significant focus in \ac{HCI}~\cite{bunian2021vins,chen2019gallery,dow2010parallel,herring2009getting,huang2019swire,lu2024flowy,ritchie_dtour_2011,mozaffari_ganspiration_2022,wu2021}.
Current research generally falls into two categories. 
The first category employs a bottom-up, serendipitous approach,  inspired by UI designer community platforms such as Mobbin~\cite{mobbin}, Dribbble~\cite{dribbble} and Behance~\cite{behance},  where designers explore a broad collection of examples to spark inspiration without a pre-defined goal~\cite{chen2019gallery, ritchie_dtour_2011}.
While this broad search space encourages discovery, it can be overwhelming and may lead to ``design drift''~\cite{mozaffari_ganspiration_2022}, where designs gradually deviate from their original goals due to ad-hoc choices during exploration.
The second category utilizes a top-down, targeted approach, where inspirational tools find examples based on specific design inputs, such as sketches or existing layouts~\cite{behrangGUIfetchSupportingApp2018,leeDesigningInteractiveExample2010,ritchie_dtour_2011,swearnginRewireInterfaceDesign2018,mozaffari_ganspiration_2022}. 
For instance, Park et al.~\cite{parkLeveragingMultimodalLLM2025} introduced a multimodal search framework that extracts semantics from input UI images and retrieves relative UI designs examples based on semantic similarity.
These UI examples offer a broad search space, requiring users to browse through many examples. 
Although targeted retrieval reduces search effort, it still relies heavily on users' ability to formulate precise inputs and offers limited support for adapting retrieved examples into personalized designs.
More recent work such as Lu et al ~\cite{luMistyUIPrototyping2024}, explores how users can blend provided examples into generative workflows, yet such systems require users to supply initial examples rather than retrieve them from large-scale repositories. 
However, this approach depends on users providing a desirable example as input, which can be particularly challenging for those who struggle to define their expectations, and may expose them to a narrow range of similar styles, leading to design fixation~\cite{mozaffari_ganspiration_2022, JANSSON19913}. 

These limitations raise a central question: 
How can we design a system that helps end users articulate their UI intentions while offering traceable justifications for design choices? 
In response to this, we propose \sys{}, a system that supports end users in designing UIs by helping them articulate their design ideas through traceable UI examples.
We refer to traceable UI examples as retrieved interfaces whose sources, context, and quality indicators are transparent to the user, allowing them to understand where a design came from and why it was recommended.
At the core of \sys{} is an example-driven exploration paradigm, where users iteratively search, select, and adapt retrieved examples to progressively shape their designs.
\sys{} leverages a \ac{MMRAG} model to retrieve UI examples based on users' text queries and their selected UI components, ranging from the whole interface to a single component.
The process begins with users generating an initial UI by simply describing their needs through a natural language query. They can then adapt this initial UI to different styles by providing another query that searches for UI examples relevant to their needs, upon which \sys{} returns a diverse set of example UIs. Users can browse these examples, select those they find satisfactory, and adapt styles, layouts, colors, or other attributes into their current design --- a process we call \textit{global remix}. 
Beyond global remix, users can also select specific components within their current design, retrieve new UI examples relevant to the selected part, and adapt styles or attributes into that component --- a process we call \textit{local remix}.
In the following iterations, users can freely use global and local remix to help design their own UI.

Drawing from social transparency theory~\cite{ehsan2021,stuart2012social}, we highlight key attributes of retrieved UI examples, such as ratings, download counts, categories and developer information, to make their sources and context visible.
This transparency enables users to assess the relevance and applicability of each design, fostering trust and critical evaluation in the system's suggestions.
While \sys{} is technically capable of generating interactive and multi-screen UIs, our work focuses on supporting the early-stage prototyping of static single-screen UI designs.

To evaluate \sys{}'s effectiveness in supporting end users during mobile UI design, we conducted a within-subjects lab study with 24 participants. 
Our analysis showed that \sys{} significantly improved participants' ability to achieve their design goals, facilitated effective iteration, and encouraged exploration of alternative designs.
Participants reported greater willingness to use \sys{} for personal UI designs, attributing these gains to the example-driven workflow and source transparency cues. 
The retrieval and presentation of real-world UI examples encouraged active exploration and inspiration, while the system's iterative refinement capabilities supported flexible, goal-oriented design improvement. 
Source transparency cues such as ratings and developer information further enhanced users' trust and justification of design decisions. 
Building on these findings, we derived several design implications for developing example-driven AI systems that enable end users to design with greater control, trust, and creativity. We also reflected on \sys{}'s design philosophy, emphasizing the role of transparent, example-driven interactions in fostering human–AI collaboration in creative design contexts.
In summary, our work makes the following contributions:


\noindent (1) \sys{}, an interactive system that employs an example-driven exploration paradigm, grounding UI designs in traceable, real-world examples to enhance intent articulation and user trust.

\noindent (2) An empirical evaluation demonstrating that \sys{} effectively supports end users in articulating their design intent and enhances users' trust in design decisions.

\section{Related Work}
\begin{table*}[t]
\centering
\caption{Summary of prior systems on multimodal retrieval, remix features, and target users. 
\checkmark indicates the feature is supported.} 

\label{tab:tab5}
\setlength{\tabcolsep}{4pt} 
\begin{tabular}{@{}l
    p{1.2cm}<{\centering}
    p{2.2cm}<{\centering}
    p{3.8cm}
    p{0.8cm}<{\centering}
    p{0.8cm}<{\centering}
    p{1.7cm}<{\centering}
    p{1.2cm}<{\centering}
@{}}
\toprule
\textbf{System} &
\multicolumn{3}{c}{\textbf{Multimodal Retrieval}} &
\multicolumn{2}{c}{\textbf{Remix}} &
\multicolumn{2}{c}{\textbf{Target User}} \\
\cmidrule(lr){2-4} \cmidrule(lr){5-6} \cmidrule(lr){7-8}
 & Keyword-based & Natural Language-based & Retrieval Type & Global & Local & Professional Designers & End User \\
\midrule
Misty~\cite{luMistyUIPrototyping2024} &  &  &  & \checkmark & \checkmark & \checkmark &  \\
Gallery D.C.~\cite{chen2019gallery} & \checkmark &  & text-to-text &  &  & \checkmark &  \\
VINS~\cite{bunian2021vins} &  &  & image-to-image &  &  & \checkmark &  \\
InspirationSearch~\cite{parkLeveragingMultimodalLLM2025} &  & \checkmark & text-to-text &  &  & \checkmark & \checkmark \\
GANSpiration~\cite{mozaffari_ganspiration_2022} &  &  & image-to-image & \checkmark & \checkmark & \checkmark &  \\
UIClip~\cite{wu_uiclip_2024} &  & \checkmark & text-to-image &  &  & \checkmark &  \\
GUing~\cite{guiing2025} &  & \checkmark & text-to-text, text-to-image, sketch-to-image &  &  & \checkmark &  \\
Swire~\cite{huang2019swire} &  &  & sketch-to-image &  &  & \checkmark &  \\
d.tour~\cite{ritchie_dtour_2011} & \checkmark &  & text-to-image &  &  & \checkmark & \checkmark \\
\rev{GUI2WiRe~\cite{10.1145/3324884.3415289}} & & \rev{\checkmark} & \rev{text-to-text} & & \rev{\checkmark} & \rev{\checkmark} & \rev{\checkmark} \\
\rev{RaWi~\cite{kolthoff_data-driven_2023}} & & \rev{\checkmark} & \rev{text-to-text} & & \rev{\checkmark} & & \rev{\checkmark} \\
\hline
\sys{} (this work) &  & \checkmark & text-to-image & \checkmark & \checkmark &  & \checkmark \\
\bottomrule
\end{tabular}
\vspace{2pt}
\small 
\end{table*}

\subsection{Using Examples in UI Designs}
Getting inspiration from existing UI examples is a common practice in UI design~\cite{bunian2021vins,chen2019gallery,dow2010parallel,herring2009getting,huang2019swire,lu2024flowy,ritchie_dtour_2011,mozaffari_ganspiration_2022,10.1145/3324884.3415289, kolthoff_data-driven_2023}. Designers frequently seek inspiration from existing designs to understand best practices, explore styles, and refine design concepts~\cite{luMistyUIPrototyping2024}. 
Gallery-based approaches such as Mobbin~\cite{mobbin}, Dribbble~\cite{dribbble}, and Behance~\cite{behance} provide large-scale visual archives that enable browsing across diverse interface styles. 
These repositories support open-ended exploration, but they often require extensive manual searching and rely on subjective interpretation.
In parallel, prior studies have explored retrieval and generation-based approaches that automatically present design examples based on user input~\cite{bunian2021vins, chen2019gallery, huang2019swire, mozaffari_ganspiration_2022, ritchie_dtour_2011}. 
Early approaches employed keyword-based retrieval~\cite{ritchie_dtour_2011,chen2019gallery}, while more recent work leverages deep generative networks and multimodal representations~\cite{wu_uiclip_2024, luMistyUIPrototyping2024,cheng2024colay,play2023,swearngin2020scout} to recommend or synthesize examples aligned with textual or visual cues. 
Other studies extend example use through style or layout transfer between UIs~\cite{beltramelli2018, AutoIcon2022, bricolage2011, luMistyUIPrototyping2024,nguyen2015, nichols2006, swearngin2018, warner2023}, allowing designers to apply desired designs from one interface to another.

While these approaches inspire users with existing examples, each category has limitations. 
The gallery-based approach offers a broad search space but can overwhelm users and lead to design drift~\cite{mozaffari_ganspiration_2022}. 
Retrieval and generative systems require users to provide clear examples or descriptions~\cite{bunian2021vins,luMistyUIPrototyping2024}, which can be particularly challenging for end users who struggle to define their expectations.
This reliance on explicit input can also lead to design fixation~\cite{mozaffari_ganspiration_2022, JANSSON19913}, as users may focus too narrowly on the few examples they can describe. 
Moreover, few systems make example sources transparent or traceable~\cite{ehsan2021, stuart2012social}, limiting users' ability to evaluate credibility and relevance of examples.




\subsection{Supporting End Users in UI Designs}
While traditional interface design tools are primarily intended for professional designers, recent research in \ac{HCI} has increasingly focused on empowering end users without formal design training to create user interfaces.  
These studies seek to lower the entry barrier to UI design by allowing users to express intent and customize the UI through accessible interactions such as direct manipulation or natural language~\cite{lee2020guicomp,kim2022,squire2025}.

Early research in this direction focused on enabling end users to restyle or extend existing interfaces through lightweight interactions. 
Systems such as Stylette~\cite{kim2022} demonstrate how users can restyle UI designs using natural language, while subsequent work such as ProgramAlly~\cite{herskovitz2024} enables end users to design and customize interfaces through multimodal instructions and demonstrations.
Additional research leverages existing UI examples to support creative exploration~\cite{parkLeveragingMultimodalLLM2025,ritchie_dtour_2011}, showing how example galleries and inspirational search can help users draw ideas and inspiration for their own designs.

Expanding on these approaches, other studies have introduced AI-assisted recommendation frameworks that guide end users during the design process~\cite{lee2020guicomp,deepinventor2021,DesignRequirements2021}.  
For example, GUIComp~\cite{lee2020guicomp} integrates with GUI design software to provide real-time, multi-faceted feedback on users' interface prototypes, helping them iteratively refine layouts and improve overall usability.
Similarly, DeepInventor~\cite{deepinventor2021} employs deep learning to convert hand-drawn interface sketches into runnable wireframes in App Inventor, supporting end users in rapidly prototyping mobile UIs.
Such systems exemplify how AI can provide guidance in UI design while preserving user control and interpretability.  

In summary, while existing works lower the barrier for end users in UI designs, they rarely support the articulation of design intent through UI examples or provide transparent justifications for design decisions.
Building on this insight, our system focuses on supporting intent articulation through traceable, example-driven design exploration and remixing.

\subsection{AI Support for UI Designs}
Recent advances in AI have transformed how designers and end users create and iterate on user interfaces.
Extensive research has examined how AI can support UI design workflows\rev{~\cite{khan2025,Knearem2023, li2024, lu2024ai, lu2022, 10.1145/3706598.3713932, 10.1093/iwcomp/iwab006}}.
Industrial systems from OpenAI, Figma, Stitch, Magic Patterns, and Vercel\footnote{Stitch: \url{https://stitch.withgoogle.com/}; Magic Patterns: \url{https://www.magicpatterns.com/}; Vercel: \url{https://vercel.com/templates/material-ui}} have further demonstrated the practical potential of AI for automating UI design tasks. However, these tools largely rely on conversational interfaces, where users must articulate their goals through precise natural language instructions, which poses challenges for end users who struggle to express abstract UI preferences.
\rev{While systems like Figma rely on similarity-based ``find more like this'' workflows that retrieve similar UIs and require manual import and remixing, our system supports preference-driven retrieval from a broader set of real-world app UIs and enables direct remixing of selected designs into the current prototype.}

In the research domain, large multimodal models have shown strong capabilities in transforming between textual, visual, and code representations of UIs\rev{~\cite{si-etal-2025-design2code,wu-2024-uicoder,luMistyUIPrototyping2024, cheng2024colay,li2023, lu2023ui, duan_uicrit_2024}}.
Research such as Design2Code~\cite{si-etal-2025-design2code} and UICoder~\cite{wu-2024-uicoder} translates UI screenshots into executable code, while others generate UI layouts from natural language descriptions~\cite{li2023, lu2023ui, le2020, li2023starcoder, yin-neubig-2017-syntactic}.
More recent works integrate multimodal reasoning for creative tasks  ---  combining layout understanding, component recognition, and style adaptation  ---  to produce coherent interface outputs~\cite{luMistyUIPrototyping2024, cheng2024colay}.
These approaches demonstrate AI's generative strength but typically operate as one-shot translators from user input to UI output, offering limited transparency into how or why a particular design is produced.

Our work builds on prior research while shifting from generation-only paradigms to retrieval-augmented, example-driven exploration.
\sys{} leverages a \ac{MMRAG} model to present real-world UI examples that help users articulate, refine, and justify their design choices.
By highlighting source transparency cues, such as ratings, developer information, and download counts,  \sys{} makes design decisions more interpretable and trustworthy.
As summarized in Table~\ref{tab:tab5}, our system specifically targets end-user designers and uniquely combines natural-language-based text-to-image multimodal retrieval with both global and local remix functions.
In doing so, our system extends the role of AI from a generative assistant to a transparent collaborator that scaffolds intent articulation through traceable, example-grounded interactions.

\section{\sys{} System Design} \label{sec:sys_design}
We introduce \sys{}, a system that supports end users in designing mobile UIs by helping them articulate their design ideas through traceable UI examples.
We first present the design motivations that guided the development of \sys{}, grounded in insights from prior literature.

\subsection{Design Motivations}

\subsubsection{Support Intent Articulation through Example-Guided Exploration}
End users struggle to articulate their design intent due to the implicit and ill-defined nature of their preferences~\cite{lyngsTellMeWhat2018, metzler2006beyond}. 
Traditional example-based UI inspiration tools either present users with a wide collection of designs~\cite{dribbble, behance} or expect them to provide a high-quality example to initiate targeted searches~\cite{behrangGUIfetchSupportingApp2018,leeDesigningInteractiveExample2010,ritchie_dtour_2011,swearnginRewireInterfaceDesign2018,mozaffari_ganspiration_2022}.
However, both strategies fall short when users cannot clearly define their needs.

This challenge highlights the need for systems that can scaffold the articulation of vague or intuitive preferences into concrete design goals.
In response, \sys{} is designed to help users gradually form and articulate their intent by iteratively searching for relevant UI examples and adapting these designs to their own UIs. The global remix supports early-stage ideation by exposing users to diverse UI examples, while the local remix enables targeted refinement of specific design aspects.
This example-driven approach helps users both articulate and evolve their design goals over time.

\subsubsection{Enable Trustworthy and Justifiable Design Choices}
Even when users are presented with suitable UI examples, they may hesitate to incorporate them into their own designs if the rationale behind those examples is unclear~\cite{hertzum2002trust}. 
Prior tools often lack mechanisms to make the source of design examples transparent~\cite{dribbble,behance}, making users question whether the suggestions are appropriate for their goals.

To address this, our design is informed by principles from social transparency theory~\cite{stuart2012social}.
Social transparency emphasizes the importance of making the intentions, identities, and actions of others visible, helping users understand the social context of information and thereby enhancing trust~\cite{stuart2012social, ehsan2021}.
\sys{} is intended to present attributes such as ratings, download counts, categories, and developer information alongside each retrieved example, allowing users to assess their relevance and traceability. This transparency is envisioned to build user confidence in design decisions and encourage meaningful adaptation of retrieved examples.

\subsubsection{\rev{Supporting Focused Exploration and Reducing Design Drift}}
Example-driven UI exploration can lead to design drift, where users lose track of their original design goals because they get distracted by browsing many unrelated examples~\cite{mozaffari_ganspiration_2022}. This phenomenon is especially common in unstructured environments like design galleries~\cite{dribbble,behance}. 
Conversely, narrowly scoped tools may induce design fixation, leading users to converge prematurely on a small set of similar solutions~\cite{JANSSON19913}.

\sys{} seeks to balance breadth and focus by offering both global and local remix functions. Global remix is designed to encourage broad inspiration, while local remix supports the targeted refinement of specific components. This structure is motivated by the need to help users stay focused on their evolving design goals \rev{during exploration}. By enabling iterative selection and blending of examples, \sys{} supports \rev{focused exploration while preserving user control over design direction.}

\subsection{\sys{} Overview}

As shown in Figure~\ref{fig:fig1}, \sys{} supports end users in designing UIs through an example-guided workflow powered by a \ac{MMRAG} model. The interface consists of three main panels. The Conversation Panel (Figure~\ref{fig:fig1}A) allows users to describe their intended interface using free-form natural language. The Example Gallery (Figure~\ref{fig:fig1}B) displays relevant UI examples retrieved based on the user's query and selected components. Each example is accompanied by metadata such as ratings, download counts, etc to help users assess its credibility. The Editable Canvas (Figure~\ref{fig:fig1}C) presents a live preview of the generated interface, where users can customize and refine their design through direct interaction with individual components.

\sys{} supports a flexible, example-driven workflow that integrates three complementary modes rather than a fixed linear sequence:
\begin{itemize}
\item \textbf{Chat Mode:}
In this mode, users can converse with our system to generate and iteratively refine their UI designs.
Through natural-language prompts, they can produce an initial complete UI or modify existing UI within the Editable Canvas.
\item \textbf{Search Mode}: In this mode, users can search for real-world UI designs using natural language queries. Retrieved designs are displayed in the Example Gallery alongside their metadata. 
If a user selects a UI element in the Editable Canvas, the system focuses retrieval on that specific component. 
\item \textbf{Apply Mode}: In this mode, users can remix retrieved UI designs into their own. They can provide natural language instructions to specify how the retrieved design should be applied to their existing UI while also highlighting elements from the selected example to indicate which parts to incorporate.
When a specific UI element on the Editable Canvas is selected, the remix operation updates only that element.
\end{itemize}

This combination of high-level inspiration and fine-grained control allows users to construct custom UIs without writing code. Users can seamlessly switch between modes, with all interactions recorded in the Conversation Panel to ensure transparency and traceability.

\subsection{Target Users and System Walkthrough}

\sys{} is designed for end user designers: non-professional creators such as small business owners, students, and hobbyists who engage in interface design as a means to an end (e.g., promoting content, prototyping ideas), rather than as their primary occupation.
This notion extends prior work in end user UI design, emphasizing the creative and visual aspects of design rather than programming.
End users often rely on UI references or existing templates but face challenges when translating their ideas into concrete UIs~\cite{lee2020guicomp}. \sys{} offers an example-driven workflow to support these users in exploring design possibilities, understanding design components, and assembling custom UIs without requiring direct manipulation of code. It assumes that users are self-directed and curious, and prefer working iteratively with visual examples to shape their creative intent.

We now walk through an example of how a user, Alice, a restaurant owner designing a mobile app for her business, might use \sys{} to create her UI:
Alice begins in the Conversation Panel (Figure~\ref{fig:fig1}A), where she enters a natural language query: \textit{``a mobile UI for a Chinese restaurant, with a menu section including dishes:...''}.
Upon submission, the system generates the corresponding code and provides a UI preview in the Editable Canvas (Figure~\ref{fig:fig1}C), where she can toggle between the preview and the generated code.

Although the current UI contains all the information she needs, Alice is not satisfied with its color scheme. She wants to see how existing restaurant UIs look, so she enters a new query in the Conversation Panel (Figure~\ref{fig:fig1}A), switches to ``search'' mode, and retrieves a gallery of real-world UI examples in the Example Gallery (Figure~\ref{fig:fig1}B), drawn from mobile UI designs that semantically align with her input.
As she browses the gallery, Alice discovers several UIs with engaging color schemes. To prioritize credible and relevant examples, she consults the Design Traceability features --- ratings, download counts, and developer information --- which provide useful signals about the quality and popularity of each UI. She selects one example, switches to ``apply'' mode in the Conversation Panel (Figure~\ref{fig:fig1}A), and specifies how she would like to adapt the current design (e.g., applying the color scheme).
Upon submission, the system updates the UI in the Editable Canvas (Figure~\ref{fig:fig1}C). We call this process a global remix.

Next, Alice decides that the reservation button design could be further improved. To refine the UI, she clicks on the button in the Editable Canvas and types her query  ---  ``a stylish red   reservation button''  ---  into the Conversation Panel, switching to ``search'' mode. The system presents a set of reservation button variations from existing UI designs. She selects a UI example featuring a rounded red button, zooms in on the design, and \textbf{marks the button she likes by drawing a circle around it}. Then, she types ``use this style'' in the Conversation Panel and switches to the ``apply'' mode to integrate the button's style into her current design. We call this process a local remix.

Alice performs several rounds of global and local remix iteratively until she is fully satisfied with her UI. 
This example illustrates how \sys{} balances creative freedom with example guidance, enabling novice users to design UIs with confidence.

\subsection{Global Remix}
\subsubsection{UI Retrieval}
Users begin by describing their intended interface in natural language through the Conversation Panel (Figure~\ref{fig:fig1}A). \sys{} then retrieves a diverse set of UI examples from a curated dataset of mobile UIs. These examples are ranked by semantic similarity to the user's query using the underlying \ac{MMRAG} model.  
The retrieved results are displayed in the Example Gallery (Figure~\ref{fig:fig1}B), where each UI example is shown alongside traceability metadata such as ratings, download counts, comment counts, and categories. These cues help users assess the credibility and relevance of each example.  
Users can also zoom in on a UI example to view a larger version, along with additional source metadata including the developer's information and app name.  

\subsubsection{UI Example Selection and Remixing}
Users browse the retrieved examples and select those that align with their design goals. 
Upon selecting an example, they can specify how they wish to adapt it, such as modifying the color scheme or layout, by entering a request in the Conversation Panel. 
After a user selects an example, the system automatically switches to ``apply'' mode.  
Once the user submits their request, \sys{} generates the corresponding HTML/CSS code and displays a live UI preview on the Editable Canvas (Figure~\ref{fig:fig1}C). 
Users can toggle between the code and the rendered preview to inspect or refine the layout. The preview is fully editable, enabling users to continue building on the layout or initiate a local remix for further customization.  

\subsection{Local Remix}
\subsubsection{UI Component Retrieval}
Local remix supports the refinement of specific components within the current UI design. 
\rev{By entering a query in the Conversation Panel (e.g., \textit{``a stylish red button''}), users trigger a component retrieval process that presents design variations of the queried component from existing UI examples. 
This retrieval shares the same backend retrieval pipeline as UI retrieval within global remix.}
The retrieved results are displayed in the Example Gallery, where each variation of the selected component is shown alongside traceability metadata.

\subsubsection{UI Example Selection and Remixing}
Users can review the retrieved component variations and select one at a time. 
\rev{When applying a retrieved variation, users may click on a corresponding UI element in the Editable Canvas to indicate where the change should be applied.}
Similar to the global workflow, they can issue natural language commands, such as ``use this color'', to adapt the chosen variation into their current design.  
Users may also zoom in on retrieved variations and highlight the specific part they wish to adapt by directly drawing on the example UI. 
Once confirmed, the system switches to ``apply'' mode in the Conversation Panel, and upon submission updates the relevant portion of the code and the UI preview accordingly.  
This fine-grained, component-level editing enables users to iterate efficiently without restarting their entire design. Multiple rounds of local refinement can be performed, and users can flexibly alternate between global and local remix as their design evolves.

\subsection{Technical Architecture}\label{sec:pipeline}

\begin{figure*}[h]
    \centering
    \includegraphics[width=\textwidth]{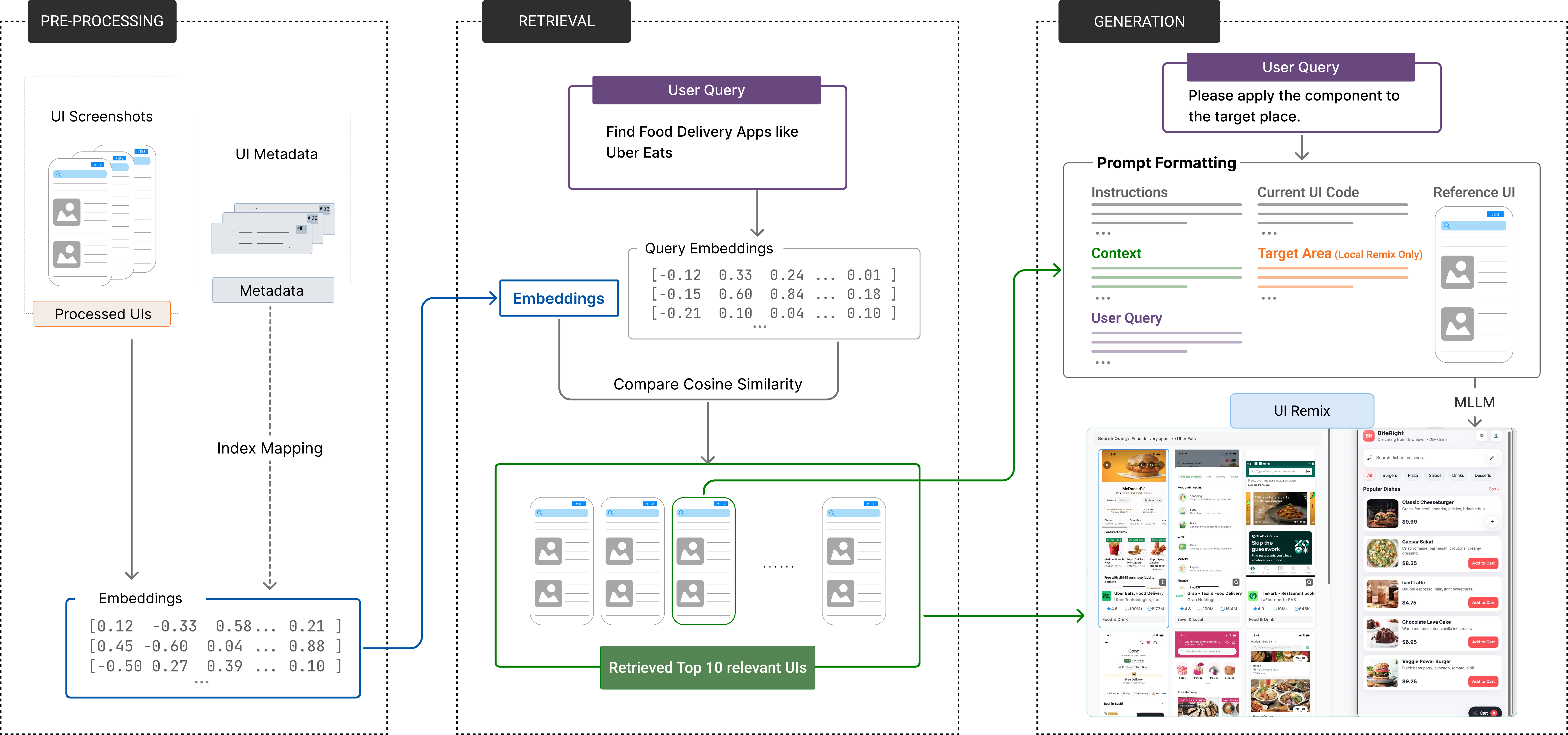}
    \vspace{-10pt}
    \caption{
\ac{MMRAG} Pipeline. The process begins with preprocessing UI screenshots from the Mobbin~\cite{mobbin}, Interaction Mining~\cite{interactionminingODIM}, and MobileViews~\cite{gao2024mobileviews} datasets, along with metadata from the Google Play Store. Each screenshot is linked to its corresponding metadata through indexing.
Next, the UI screenshots are passed through an encoder to generate embeddings. The embeddings, together with the corresponding image content, are stored in ChromaDB. When a user submits a query, it is converted into embeddings, and the system retrieves and ranks relevant screenshots using cosine similarity.
The retrieved UIs, along with their metadata, are then displayed in the Example Gallery for the user to browse and select desirable designs.
Once a user selects a UI, the selected design --- combined with the user's query, current UI code, and target area (used only for local remix) --- is formatted into a complete prompt and sent to the \ac{MLLM}, which generates updated UI code in a diff-patch format. The system backend parses these responses using the \texttt{unidiff} and \texttt{difflib} packages, updates the UI code, and renders them as UI previews.}
    \label{fig:fig2}
\end{figure*}

\sys{} is a UI design system powered primarily by a \ac{MMRAG} framework.
As illustrated in Figure~\ref{fig:fig2}, \sys{} is designed to support three modes of interaction (Chat, Search, Apply) through two core modules: the retriever and the generator. This section introduces the data source utilized in the pipeline and provides a brief overview of both modules.

\subsubsection{Data Source and Preprocessing}
Our data sources are the Mobbin~\cite{mobbin}, Interaction Mining~\cite{interactionminingODIM} and the MobileViews~\cite{gao2024mobileviews} dataset. The Mobbin is an online UI screenshot browsing platform that provides UI screenshots of mobile apps and websites. Interaction Mining is a online platform for managing and sharing mobile interaction dataset, which provides UI screenshots of the track implementing various operations. MobileView is a large-scale dataset designed for research on mobile UI analysis and mobile agents, which contains more than 600K UI screenshot-view hierarchy pairs from more than 20,000 applications. The metadata of the applications are manually collected from the Google Play Store.

We selected 196 popular mobile applications across diverse categories, including Food, Travel, Lifestyle, News, Education, etc. Subsequently, we collected corresponding UI screenshots from the Mobbin, Interaction Mining, and MobileViews datasets, yielding around 900 unique interface screenshots that served as the example set to support users during the design process.
Although we did not collect UI data for all app categories due to resource constraints, our pipeline can easily scale to larger datasets by providing screenshots and corresponding metadata.
We then processed the collected screenshots by cropping them to eliminate black borders and ensure consistent framing across all images.

\subsubsection{Retriever and Generator Modules}  
To enable example retrieval, we construct a UI screenshot database by embedding each UI screenshot into a vector representation using GUIClip~\cite{guiing2025}. The resulting embeddings are indexed and stored in a ChromaDB vector database~\cite{chromadb24}, allowing efficient similarity-based retrieval during interaction.

When users switch between modes, the system handles them differently.

\begin{itemize}
    \item \textbf{Chat Mode:} In this mode, the user's query is combined with the current UI code (which is empty on the first call) to form a structured prompt that is processed by the \ac{MLLM}. To reduce generation time, the \ac{MLLM} is instructed to produce code in a diff-patch format, representing only the changes needed. The generated patch is then applied to the existing UI code to update the design accordingly.
    \item \textbf{Search Mode:} \rev{In this mode, the system first encodes the user's natural language query into a fixed-dimensional vector representation using GUIClip~\cite{guiing2025}, which maps textual descriptions into the same embedding space as UI screenshots. It then computes the cosine similarity between the query vector and each stored screenshot embedding, defined as the normalized dot product between the two vectors, to measure their semantic closeness.} Based on these similarity scores, the system retrieves the top 10 most relevant screenshots\footnote{The choice of retrieving the top 10 results follows common practice in information retrieval~\cite{jones1999phrasier}.}. These retrieved UI designs are ranked by similarity and presented in the Example Gallery, accompanied by their associated metadata.
    \item \textbf{Apply Mode:} In this mode, the system supports two remix operations --- global and local --- depending on the scope of the user's modification request.For global remix, the system combines the user's natural language query, the current UI code, and the selected reference screenshot into a structured prompt. This prompt is sent to the GPT-5 model~\cite{openaiIntroducingGPT5}, which generates an updated version of the entire UI code reflecting the requested changes. For local remix, the system constructs a structured prompt that includes the user's query, the current UI code, the reference screenshot annotated with the user's drawing, and the identifier of the target UI component. The GPT-5 model then generates code in a diff-patch format (consistent with the format used in Chat Mode) as only a specific region of the interface is modified. The system backend parses the diff-patch codes using the \texttt{unidiff}~\cite{githubGitHubMatiasbpythonunidiff} and \texttt{difflib}~\cite{pythonDifflibHelpers} packages, updates the UI code, and renders the updated interface at the Editable Canvas.
\end{itemize}

\subsubsection{Implementation}\label{sec:implementation}
We implemented \sys{} as a web-based application for easy access.
The backend, built with FastAPI, integrates the full generation and retrieval pipeline. It uses the OpenAI API to send requests to the GPT-5 model~\cite{openaiIntroducingGPT5} and ChromaDB~\cite{chromadb24} to store and query screenshot embeddings for retrieval. We selected GPT-5 because our task focuses on general UI design code generation; at the time of the study, GPT-5 was among the best-performing models for code generation~\cite{openaiIntroducingGPT5} and was therefore considered appropriate for our setting.

The generated UI designs are rendered on the frontend within an iframe element~\cite{w3HTMLStandard}. 
To support interactive remixing, we inject custom CSS and JavaScript into the iframe, enabling users to select components for local remix directly within the preview panel. 
The backend communicates bidirectionally with the frontend --- receiving user actions and returning model outputs along with retrieved UI examples.

The frontend was developed from scratch using React. The iframe~\cite{w3HTMLStandard} is responsible for rendering the generated UI designs, while the Monaco Editor~\cite{microsoftMonacoEditor} provides a built-in code editor that allows users to view and modify UI code.
\footnote{\rev{The pipeline is open-sourced at: https://github.com/ETH-PEACH-Lab/UI\_Remix.}}

\subsubsection{Pilot Studies and Iterative Refinement}\label{sec:pilot}
We conducted four pilot studies with four participants: three master's students in data science and one PhD student in computer science. The goal was to collect feedback on the system's design and usability.
\rev{In these pilot sessions, participants completed the same UI design tasks as those used in the main study (see Section~\ref{sec:user_study}), using an early version of the system. Each session lasted approximately 20--30 minutes and focused on identifying usability issues and interaction breakdowns. Feedback was captured through researcher observation notes and informal post-task interviews.}

A key issue raised during the pilot sessions concerned the long waiting time for revising UI designs. In the earlier system version, the \ac{MLLM} was prompted to regenerate the entire UI code even for small revisions, which took over three minutes on average. To address this, we modified the prompt to request only \emph{diff-patch} code and implemented an update mechanism using the \texttt{unidiff} and \texttt{difflib} packages. This optimization reduced the average revision time to about one minute.

Another issue was the lack of a mechanism to revert to previous design versions after making edits. To address this, we implemented a version history system that records all intermediate design states. Users can now navigate between versions using the ``Back'' and ``Forward'' buttons, allowing them to easily revisit or restore any earlier version of their design.
Further improvements included repositioning the \say{Home} button and refining the interface for switching between the ``Chat'', ``Search'', and ``Apply'' modes.

\subsubsection{Technical Evaluation}
\begin{table}[h]
    \centering
     \caption{Retrieval performance of the system across different query types. Hit@5 measures the proportion of queries retrieving at least one relevant result in the top 5, while nDCG@5 reflects the quality of ranking among the top results.}
    \begin{tabular}{p{3.3cm}p{1.5cm}p{1.8cm}}
     \toprule
      \textbf{Query Type} & \textbf{Hit@5} & \textbf{nDCG@5} \\
      \midrule
      Color Theme & 0.92 & 0.83 \\
      Layout & 0.88 & 0.72 \\
      UI Category & 0.92 & 0.84 \\
      UI Component & 0.80 & 0.70 \\
      \midrule
      \textbf{Average (All)} & \textbf{0.88} & \textbf{0.77} \\
      \bottomrule
    \end{tabular}
   
    \label{tab:tab4}
\end{table}
To assess the retrieval performance of our \ac{MLLM}-based system, we conducted a controlled technical evaluation using 100 automatically generated text queries. The queries were created from predefined templates covering four intent types --- \emph{color theme}, \emph{layout}, \emph{UI category}, and \emph{UI component design} --- each targeting a distinct aspect of user interface representation. This setup allows us to systematically test whether the retriever can align textual descriptions with corresponding visual or structural features in the corpus.

For each query, the system retrieved the top-5 results. Two researchers independently annotated the relevance of each retrieved UI on a 4-point scale (0 = irrelevant, 1 = slightly related, 2 = moderately relevant, 3 = highly relevant). Scores of 2 or above were considered relevant when computing \textit{Hit@5}, while the full graded values were used for \textit{nDCG@5}.
The annotators initially disagreed on seven cases out of the total, which were resolved through discussion to reach a final consensus. 

As shown in Table~\ref{tab:tab4}, our system achieved an average Hit@5 of 0.88 and nDCG@5 of 0.77 across all query types. 
\rev{
This result indicates that in 88\% of queries, at least one UI judged relevant by annotators appeared within the top five retrieved results, while the nDCG@5 score suggests that more relevant UIs were generally ranked higher in the list.
These scores are considered strong for vision-language UI retrieval tasks~\cite{guiing2025,faysse2025colpali}.
}
The retriever performed particularly well on color and category queries, demonstrating that the multimodal embeddings effectively capture both stylistic and semantic aspects of UI design. These results confirm that the retrieval module provides accurate and consistent matches, supporting the interactive exploration features evaluated in our user study.
\section{User Study} \label{sec:user_study}
To investigate the effectiveness of \sys{} in supporting end user in UI design, we conducted a user study to address the following research question:
\begin{itemize}
    \item[] \textbf{RQ.} \textit{How does \sys{} support end users in achieving their design goals, iterating on designs, exploring alternative designs, and influencing their trust through traceable UI examples? }
\end{itemize}

We employed a within-subjects study with two experimental conditions, involving 24 end users who had no expertise in mobile UI design. 
Each participant was asked to design two mobile UIs, one with each system.
Although \sys{} is capable of generating interactive and multi-screen UIs, our evaluation focused on static single-screen UI prototyping to examine UI design within the limited study duration.
This scope reflects the early stage of UI prototyping, when users focus on exploring and refining design ideas before adding interactivity.
\rev{Accordingly, our study design emphasizes divergent exploration and early ideation; although \sys{} supports iterative refinement, the limited task duration and single-screen setup did not allow for sustained or deep refinement, which we leave to future investigation.}

\subsection{Two Experimental Conditions}\label{sec:three_design}
To understand how end users perceive the benefits of \sys{}, we implemented two system variants for comparison, using the same study apparatus.
During the study, we referred to the conditions by their names (Alpha, Beta) when introducing the session to participants to minimize any bias associated with the names. 

\begin{itemize}
    \item \textbf{GPT-Canvas Baseline (Alpha)}: The GPT-Canvas baseline adopts a design similar to GPT Canvas~\cite{openaiIntroducingCanvas} and Figma Make~\cite{figmaIntroducingFigma}. It uses the same \ac{MLLM} as \sys{}, but excludes the Example Gallery (Figure~\ref{fig:fig1}B) as well as the corresponding search and apply modes in the Conversation Panel (Figure~\ref{fig:fig1}A). By omitting these components, we can evaluate whether their presence influences users' performance in mobile UI design and their perceptions of the system. \rev{Participants interacted with the system using text-only instructions.}
    \item \textbf{\sys{} (Beta)}: \sys{} incorporates all design features and functionalities described in Section~\ref{sec:sys_design}. It integrates our specially designed \ac{MMRAG} model for UI example retrieval and employs the \ac{MLLM} to generate initial UIs and perform example-based remixing.
\end{itemize}

\subsection{Participants and Recruitment}
We conducted a power analysis using G*Power \cite{faul2009statistical} to estimate the required sample size for our within-subjects study design. 
Based on pilot studies (Section~\ref{sec:pilot}), we assumed an effect size of $f = 0.6$ (moderate), with a significance level of $\alpha = 0.05$ and power ($1 - \beta$) of 0.8, resulting in a required sample size of 24 participants.
After securing ethical approval, we recruited 24 participants through social media platforms. 
\rev{Participants included 19 male and 5 female individuals, with an average age of 25.}
Qualified participants self-identified as having little or no expertise in mobile UI design but expressed interest in it.
\rev{Importantly, none of the participants were professional UI designers; we define ``mobile UI design expertise'' as formal training or professional experience in mobile UI design, whereas reported UI design experience reflects informal, non-professional activities (e.g., designing a personal website).}
Our participants varied in their familiarity with AI tools for interface design, ranging from those who had never used such tools to those who engaged with them a few times a week.
Table~\ref{tab:tab2} shows a detailed breakdown of participant demographics. 

\begin{table*}[h]
    \centering
    \scriptsize
    \caption{Participants' demographics: \rev{all 24 participants reported having \textbf{no mobile UI design expertise}.} The sample included 10 bachelor, 12 master's, and two Ph.D. degree holders. Their experience with AI ranged from 6 months to over 3 years, while their usage of AI for UI design ranged from never to a few times a week.}
    \begin{tabular}{c p{1.2cm} p{2.4cm} p{2.8cm} p{2cm} p{4.3cm}}
     \toprule
      \textbf{PID} & \textbf{Educational Level} & \rev{\textbf{Occupation}} & \textbf{Frequency of Using AI for UI Design} & \textbf{Experience of Using AI} & \textbf{Design Tools Used Before} \\
      \midrule
      1  & Bachelor & \rev{Student} & Never & 2--3 years & PowerPoint, Google Slides \\
      2  & Bachelor & \rev{Student} & Never & 2--3 years & PowerPoint, Google Slides, ChatGPT, Photoshop \\
      3  & Bachelor & \rev{Pilot/Union Representative} & Less than once a month & 1--2 years & MidJourney \\
      4  & Master & \rev{Student} & Never & 1--2 years & PowerPoint, Google Slides, Canva, ChatGPT, Photoshop, MidJourney, DALL·E \\
      5  & Master & \rev{Scientific Assistant} & Less than once a month & 1--2 years & None \\
      6  & Master & \rev{Student} & About once a week & 2--3 years & ChatGPT, Figma \\
      7  & Master & \rev{PhD Student} & Less than once a month & 2--3 years & PowerPoint, Google Slides, ChatGPT, Figma, Photoshop, DALL·E \\
      8  & Bachelor & \rev{Teacher} & A few times a month & 0.5--1 year & Google Slides, ChatGPT \\
      9  & Bachelor & \rev{Student} & A few times a week & 2--3 years & PowerPoint, Google Slides, ChatGPT \\
      10 & Master & \rev{PhD Student} & Never & 1--2 years & Google Slides, Canva, Adobe Illustrator, Photoshop \\
      11 & Master & \rev{PhD Student} & Less than once a month & 1--2 years & PowerPoint, Google Slides \\
      12 & Bachelor & \rev{Student} & Never & 2--3 years & PowerPoint, Google Slides, Canva, ChatGPT, Adobe Illustrator, MidJourney, DALL·E, MS Paint \\
      13 & Master & \rev{PhD Student} & Less than once a month & 2--3 years & PowerPoint, Google Slides, Canva, ChatGPT \\
      14 & Master & \rev{Data Scientist} & A few times a month & 1--2 years & PowerPoint, Google Slides, ChatGPT, Figma, Adobe Illustrator, Photoshop, DALL·E \\
      15 & Bachelor & \rev{Student} & A few times a week & 1--2 years & PowerPoint, Google Slides, Canva, Adobe Illustrator \\
      16 & Master & \rev{PhD Student} & A few times a month & 2--3 years & PowerPoint, Google Slides, ChatGPT, Photoshop, Others \\
      17 & Master & \rev{PhD Student} & Less than once a month & 2--3 years & PowerPoint, ChatGPT, Figma \\
      18 & Bachelor & \rev{Student} & A few times a week & 1--2 years & PowerPoint, Google Slides, Canva, ChatGPT \\
      19 & Master & \rev{Student} & A few times a week & 1--2 years & Google Slides, ChatGPT, Photoshop \\
      20 & Master & \rev{Doctorate Researcher} & A few times a month & > 3 years & PowerPoint, ChatGPT \\
      21 & Ph.D & \rev{Scientific Researcher} & Less than once a month & > 3 years & PowerPoint, Google Slides, ChatGPT, Figma, MidJourney, DALL·E \\
      22 & Ph.D & \rev{PostDoc} & Never & > 3 years & PowerPoint, Google Slides, Canva \\
      23 & Bachelor & \rev{Student} & Less than once a month & 1--2 years & PowerPoint, Google Slides, ChatGPT, Photoshop, MidJourney \\
      24 & Bachelor & \rev{Student} & Less than once a month & 2--3 years & PowerPoint, Google Slides, Canva, ChatGPT, Figma, MidJourney, DALL·E \\
      \bottomrule
    \end{tabular}
    \label{tab:tab2}
\end{table*}
\subsection{Study Protocol}
All participants signed the consent form before the study. During the study, participants first received a 5-minute introduction on how to use \sys{}. 
To ensure participants fully understood how to use each system, we demonstrated its functions and then gave them time to explore it on their own before starting each task.
We began each task only after participants indicated that they were comfortable with the system.
During this familiarization phase, participants were encouraged to ask any questions they had about using \sys{}.

\rev{Participants then completed two mobile UI design tasks, with a maximum time limit of 25 minutes per task, each representing a typical challenge in mobile UI design.}
For each task, participants designed one mobile UI using one of the two systems.
We used a Balanced Latin Square design~\cite{bradley1958complete} to arrange the order of the assisting methods for each participant, \rev{ensuring that each method was used across different tasks rather than being consistently paired with the same task}, while minimizing potential order effects.

After each task, participants completed a Likert-scale post-task questionnaire to evaluate the system. 
After completing both tasks, we conducted a 10-minute semi-structured exit interview with each participant and collected their completed UI design for analysis.

The study lasted 60-70 minutes and was conducted virtually via Zoom. 
Participants used the system in their own browsers.
Each received 25 USD as compensation.

\subsection{Study Task}
We designed two tasks to represent common mobile UI design scenarios for end users.
The first involved creating a menu page for a hamburger restaurant based on a provided menu.
The second involved designing a news display page for a news-reading app using a given set of news articles.
We selected these two tasks for their straightforward and familiar nature, which allows participants to focus on design interaction rather than domain knowledge.
We set a 25-minute time limit per task, based on pilot results showing participants could complete the tasks within that time.

\subsection{Data Collection and Analysis}
This study collected data from multiple sources, including screening questionnaires, post-task questionnaires, exit interviews, observation notes, system usage logs and expert ratings. 
We report the significance levels ($p$) and test statistics ($Z$ or $t$) for each statistical test in this section and Section~\ref{sec:result}.

\subsubsection{Post-Task Questionnaires}

The post-task questionnaires used 7-point Likert scales to evaluate systems on various attributes.
The Shapiro–Wilk test showed none of the post-task questionnaire scores were normally distributed. 
As Beta consistently had a higher mean score than Alpha condition, we performed one-sided post-hoc Wilcoxon signed-rank tests with Bonferroni correction to compare Alpha with Beta. Results are in Table~\ref{tab:tab1}.

\subsubsection{Semi-structured Interview}
We conducted exit interviews after the session, recording 329 minutes of video to gather additional feedback on the systems.
We transcribed and proofread participants' answers and the recordings to produce the interview transcripts. We use the interview transcripts as anecdotal evidence to support our quantitative findings. Therefore, we used \textit{in vivo}~\cite{saldana_coding_2013} coding to analyze the interviews and attune ourselves to the users' perspectives on the different designs.

\subsubsection{Observation Notes and Usage Logs}
During each session, a researcher took observational notes while the system recorded participants' queries, UI designs, and function usage. 
These data were used to capture user behavior and provide insights into how participants interacted with the different systems. The screen recordings were also transcribed and used to review participant behaviors. 

\subsubsection{Task Completion Time and Expert Evaluation}

Regarding task completion time, the Shapiro–Wilk test indicated no significant deviation from normality. Therefore, we conducted paired t-tests with Bonferroni correction to compare completion times between Alpha and Beta, as well as between the two tasks. The results showed significant difference in completion time between the two systems ($t$ = -3.05, $p$ < 0.01) and no significant effect of task type on completion time ($t$ = 0.45, $p$ = 0.66). 
These results are reasonable, as the Beta system includes additional functions such as global remix and local remix, which may require additional time to operate. However, the average completion time for the Beta system (15.76 min) was only approximately 3 minutes longer than that for Alpha (12.41 min), suggesting that the increased time demand was modest.

To evaluate the quality of participants' designed UIs, we collected 48 UIs and presented them in random order to two external UI design experts who were not involved in the project.
The experts jointly reviewed each design, discussed their judgments, and reached a consensus rating on three criteria --- aesthetic quality, layout quality, and accuracy --- using a seven-point Likert scale (1 = very poor, 7 = excellent).
We then conducted two-sided Wilcoxon signed-rank tests on the consensus ratings, and no significant differences were observed between the Alpha and Beta conditions across any of the criteria.
\rev{One plausible explanation is that the expert evaluation rubric is necessarily coarse-grained and focuses on static end results, which may not capture more subtle differences related to design intent or rationale.
As a result, the benefits of example-driven interaction, such as supporting how users reason about and explore design intent during the task, may not accumulate into observable differences in final UI quality within short, time-constrained tasks, and increased exploration can come at the expense of final visual polish.}
\section{Results}\label{sec:result}
We report results from the user study and summarize key findings as below.
\newcommand{\figurewidth}{5cm}
\begin{table*}[h]
\centering
\caption{Results of post-task questionnaire comparing \textit{Alpha} and \textit{Beta}. Wilcoxon signed-rank tests with Bonferroni correction were conducted for each statement. Reported are Bonferroni-corrected p-values ($p<.05$*; $p<.01$**).}
\begin{tabular}{p{5cm} p{1.2cm} p{0.3cm} p{0.3cm} p{0.3cm} p{0.8cm} c}
\toprule
\textbf{Statement} & \textbf{Condition} & \textbf{N} & \textbf{M} & \textbf{SD} & \textit{p} & \textbf{Agreement: 1 to 7} \\
\midrule

\multirow{2}{5cm}{The system is generally easy to use.} 
& Alpha & 24 & 5.75 & 1.22 & \multirow{2}{*}{0.36} & \multirow{2}{*}{\includegraphics[width=\figurewidth]{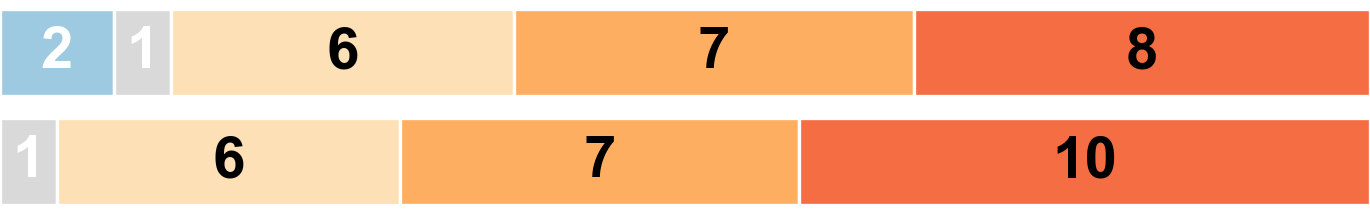}} \\
& Beta  & 24 & 6.08 & 0.93 & & \\

\midrule
\multirow{2}{5cm}{The system's interaction is intuitive.} 
& Alpha & 24 & 5.62 & 1.35 & \multirow{2}{*}{0.22} & \multirow{2}{*}{\includegraphics[width=\figurewidth]{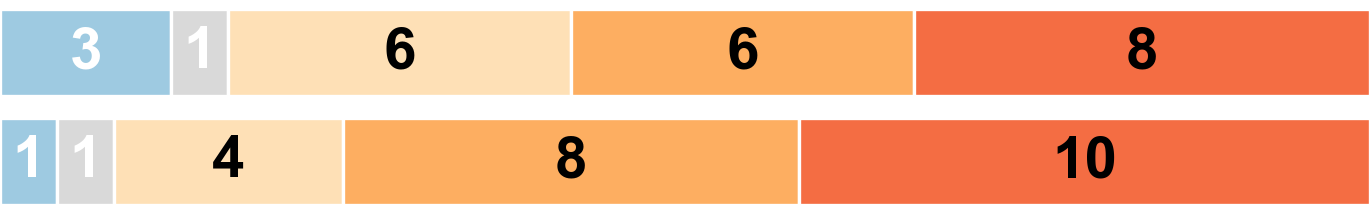}} \\
& Beta  & 24 & 6.04 & 1.08 & & \\

\midrule
\multirow{2}{5cm}{The system supports me well in achieving my design goals.} 
& Alpha & 24 & 4.79 & 1.67 & \multirow{2}{*}{\textbf{0.02*}} & \multirow{2}{*}{\includegraphics[width=\figurewidth]{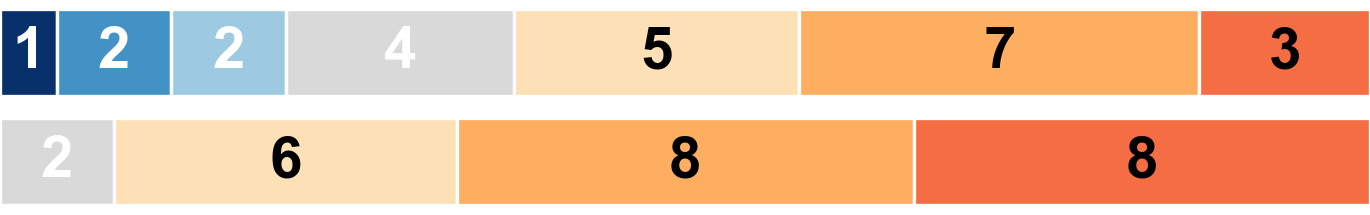}} \\
& Beta  & 24 & 5.92 & 0.97 & & \\

\midrule
\multirow{2}{5cm}{The system's outputs within each iteration follow my expectations.} 
& Alpha & 24 & 4.83 & 1.61 & \multirow{2}{*}{0.14} & \multirow{2}{*}{\includegraphics[width=\figurewidth]{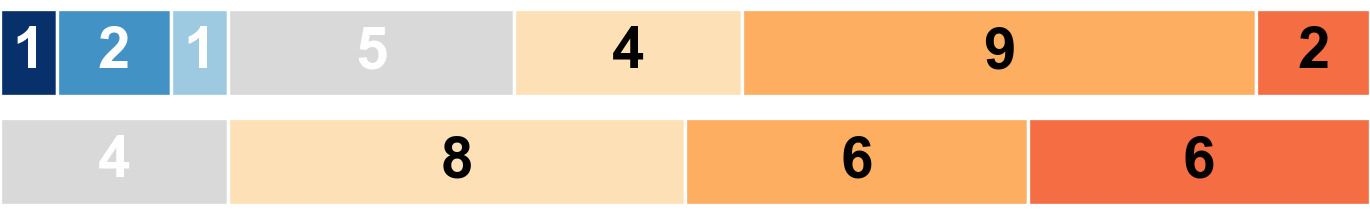}} \\
& Beta  & 24 & 5.58 & 1.06 & & \\

\midrule
\multirow{2}{5cm}{I am satisfied with the design outcomes I get in the end.} 
& Alpha & 24 & 5.33 & 1.52 & \multirow{2}{*}{0.13} & \multirow{2}{*}{\includegraphics[width=\figurewidth]{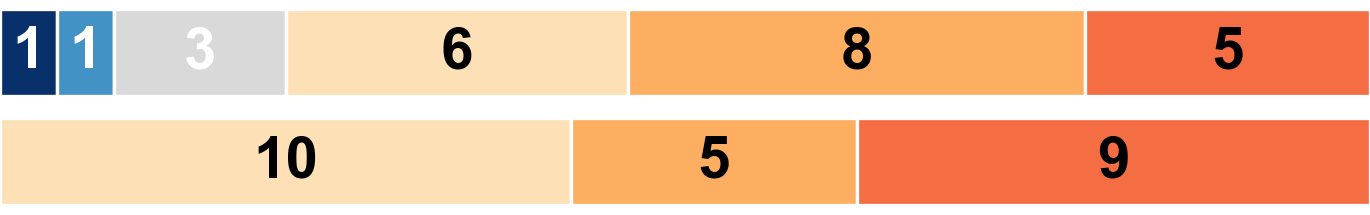}} \\
& Beta  & 24 & 5.96 & 0.91 & & \\

\midrule
\multirow{2}{5cm}{The system can help me to iterate on UI designs effectively.} 
& Alpha & 24 & 4.67 & 1.99 & \multirow{2}{*}{\textbf{0.04*}} & \multirow{2}{*}{\includegraphics[width=\figurewidth]{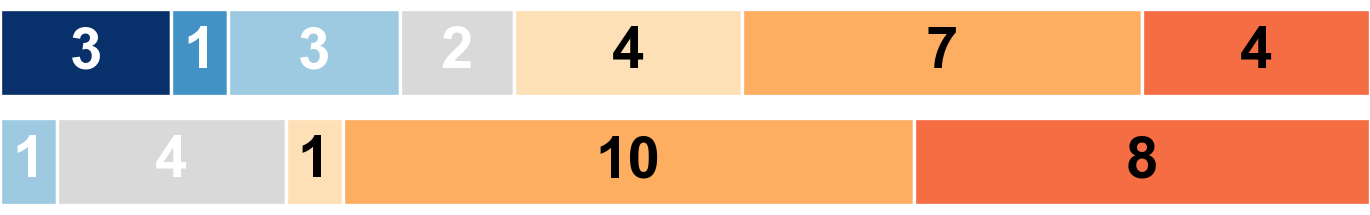}} \\
& Beta  & 24 & 5.83 & 1.20 & & \\

\midrule
\multirow{2}{5cm}{The system helps me to explore more alternative design options.} 
& Alpha & 24 & 4.17 & 2.01 & \multirow{2}{*}{\textbf{0.00**}} & \multirow{2}{*}{\includegraphics[width=\figurewidth]{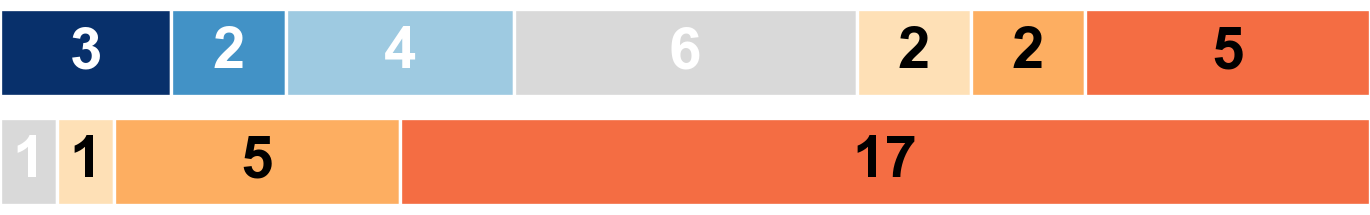}} \\
& Beta  & 24 & 6.58 & 0.78 & & \\

\midrule
\multirow{2}{5cm}{If I need to build a personal mobile UI, I would like to use this system.} 
& Alpha & 24 & 4.88 & 1.83 & \multirow{2}{*}{\textbf{0.01**}} & \multirow{2}{*}{\includegraphics[width=\figurewidth]{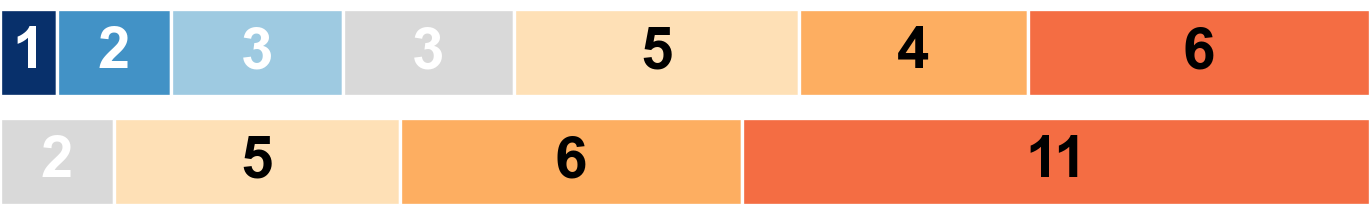}} \\
& Beta  & 24 & 6.08 & 1.02 & & \\

\midrule
\multirow{2}{5cm}{I would like to use this system for other creative tasks.} 
& Alpha & 24 & 4.75 & 1.80 & \multirow{2}{*}{\textbf{0.00**}} & \multirow{2}{*}{\includegraphics[width=\figurewidth]{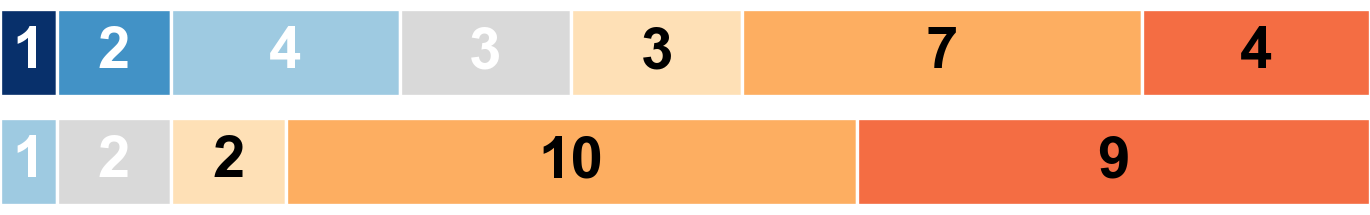}} \\
& Beta  & 24 & 6.00 & 1.10 & & \\

\bottomrule
\end{tabular}

\label{tab:tab1}
\end{table*}

\subsection{Encouraging Creative Exploration through Examples}
Participants reported that the Beta system substantially enhanced their ability to engage in creative exploration compared to Alpha.
Quantitatively, Beta received significantly higher ratings for ``helping to explore more alternative design options'' ($p$ < .001, $Z$ = 1.03) and ``like to use this system for other creative tasks.'' ($p$ < .001, $Z$ = 0.03), as shown in Table~\ref{tab:tab1}.

A central way Beta supported creativity was through inspiration from examples.
Nine participants (P1–2, P4–5, P6–7, P9–P11) explicitly mentioned that they liked the example-driven nature of the Beta system.
Participants described that seeing real-world UI examples supported their design process in three key ways. First, it helped them spark creative design inspirations (P1, P3, P10, P12); for instance, P3 mentioned that they liked to \inlinequote{get design ideas ... see inspirations and try to use them from other websites or other apps.} Second, it guided them in applying existing styles to their own designs (P2, P10, P14, P15). As P14 pointed out \inlinequote{[the retrieved examples] helped a lot; I can refer to many other different websites, and integrate their ideas into my own design.}. Finally, it enabled them to learn from established best practices (P9, P10).
These UI examples served as concrete visual references that helped participants imagine new design directions or adapt effective elements.
As P7 noted, the system \inlinequote{helps go beyond language description while building the app}, allowing them to move from abstract intentions to tangible visual ideas.

Examples also encouraged participants to explore more design options and experiment with variations.
Five participants (P4-7, P17) described that seeing multiple examples motivated them to look for ``other layouts'' or try ``more options'' before deciding on a final design.
As P7 put it, the system \inlinequote{encourages you to look for other layouts}, suggesting that the example presentation expanded users' exploration space and prompted more creative comparison among different possibilities.
This active engagement with alternative examples enabled participants to experiment freely and evaluate designs from multiple perspectives, fostering a more open design mindset.
Overall, Beta transformed example viewing from a passive experience into a more dynamic process of exploration, which allowed participants to question and imagine what their ideal UI could look like.

\subsection{Supporting Goal-Oriented Design Iteration}
Beyond fostering creative exploration, participants also highlighted how the system supported goal-oriented iteration during the design process, helping them refine their design toward their intended design goals.
Quantitatively, participants rated the Beta system significantly higher for ``helping to iterate on UI designs effectively'' ($p$ = .04, $Z$ = 0.67) and ``supporting me well in achieving my design goals'' ($p$ = .02, $Z$ = -0.64).
\rev{Behavioral data from usage logs and observation notes further support these findings. Participants used the search mode an average of 3.75 times, the global remix 1.33 times, and the local remix about 3 times per task; six participants (P3, P6–7, P11–12, P20) retrieved at least five relevant UIs, reflecting active engagement in iterative refinement.
In addition, interaction logs show that 21 participants (P1–7, P9–14, P17–24) used the retrieval mode at least once within their first five operations, suggesting that participants often referred to existing examples early in the design process.}

Participants found that examples served as reference points for meaningful iteration throughout the design process.
Five participants (P2, P6, P11, P13, P14) mentioned that the presence of concrete examples made it easier to start the design and iteration process.
As P2 explained, \inlinequote{I like that it can give some examples of the app, because I don't know what the layout will look good}, while P11 shared that the system \inlinequote{makes me feel good or easier than starting from scratch.}
These reflections suggest that examples served as visual scaffolds that guided users' goal-driven refinement.
Instead of starting over or relying on trial and error, participants could use examples to recognize what aspects worked or needed adjustment, making iteration a more guided and flexible process toward their desired outcomes.

Participants also discussed how specific system features, particularly the global remix and local remix, facilitated iterative refinement during the design process.
The local remix function was appreciated for allowing users to modify a specific component while preserving the overall design, enabling them to incorporate elements from UI examples into targeted components of their UI design (P1-2, P5). 
In contrast, the global remix supported higher-level design adjustments by applying UI styles from retrieved examples to the entire interface. 
Seven participants explicitly stated that the global remix function effectively helped them apply UI styles from selected examples (P1-2, P5, P11, P13, P15, P24), e.g., P1 \inlinequote{I would say it's actually quite powerful, since you can perfectly copy the style and layout and apply the reference I choose to our app.}


\subsection{Fostering Trust and Confidence through Source Transparency Cues}

Participants rated the Beta system significantly higher for ``If I need to build a personal mobile UI, I would like to use this system'' ($p$ = .01, $Z$ = -0.93), indicating a strong willingness to adopt it for real design tasks.
Interview data further suggest that this preference was closely and consistently tied to the source transparency cues displayed alongside retrieved UI examples.
These cues included app ratings, download counts, developer information, app name, and category, allowing users to clearly trace where each example originated.

Out of 24 participants, a clear majority of 17 actively examined these cues during the study itself. 
Participants described source transparency cues as valuable for assessing the credibility and relevance of design examples overall.
Six participants (P2-5, P10-11) mentioned that higher ratings made them more inclined to trust and adopt certain designs, reflecting their belief that well-rated UIs were more likely to come from successful or professional applications.
Others emphasized that these cues increased reliability (P1, P3) and provided additional references (P4-5) for evaluating UI examples.
As P9 noted, \inlinequote{by looking at the installation number and the rating. If these numbers are high, the user will say, okay, this may be a good design and choose this one. So the metadata serves as a reference.}
These comments indicate that source transparency cues provided an extra layer of contextual trust, enabling participants to evaluate examples not only by appearance but also by perceived credibility, making their design choices feel better informed and thus justified.

However, not all participants relied on these cues.
Seven participants (P2, P6, P12-14, P20, P22) stated that metadata did not significantly influence their design decisions, with some focusing purely on the visual aspects of UI examples.
As P6 explained, \inlinequote{I just don't really care about the details here}, and P12 mentioned, \inlinequote{I focus more on the visual aspect, so I wasn't really influenced by the metadata at all.}
This variation suggests that while transparency cues were effective in fostering trust and confidence for many users, others prioritized direct visual appeal or personal aesthetic judgment.
Overall, the findings indicate that transparency cues help build trust, particularly among users who value contextual validation, yet still allowing flexibility for those who rely more on visual intuition.
\section{Discussion and Future Work}

\subsection{Significance of Example-Driven Design in Articulating Intent and Building Trust}
\rev{Our findings suggest that, from participants' perspectives, designing with real-world UI examples can support the process of articulating design goals and developing trust in AI-assisted design.}
This process occurs through two mechanisms: (1) grounding users' underspecified intents in concrete design references, and (2) enhancing user trust and design confidence through source transparency cues.

\rev{Firstly, participants experienced the example-driven paradigm as supporting intent articulation through externalization.}
Participants often started with broad, underspecified goals such as ``a modern, clean restaurant menu'' but struggled to define what ``modern'' or ``clean'' meant in design terms. 
\rev{By browsing and comparing examples, participants reported gradually refining their design goals, using retrieved UI designs as concrete references to evaluate and adjust their preferences.}
This process often prompted self-explanations (e.g., \textit{``I prefer this one because the layout is simpler''}), \rev{suggesting that examples helped participants reflect on and verbalize preferences that had previously been implicit.}
\rev{This pattern can be interpreted through the notion of reflection-in-action: where designers come to understand their goals through interaction with concrete representations~\cite{schon2017reflective}.}

Secondly, the use of source transparency cues for UI examples directly contributed to the perceived trust and design confidence. 
Displaying source transparency cues, including ratings, developer information, and download counts, helped users evaluate the credibility of retrieved designs.
In line with social transparency theory~\cite{ehsan2021,stuart2012social}, these cues allowed users to form transitive trust, reasoning that ``if many others used or liked this design, it must be reliable.'' 
This was particularly important for novices, who lacked professional heuristics for judging design quality. 
Transparency thus transformed retrieval results from opaque outputs into ``trustworthy design references'', enabling users to make informed adaptation decisions.

\subsection{Supporting Progressive Articulation Through Iterative Co-Evolution}
A central insight from our study is that \sys{} supported users in iteratively refining their designs through interaction with examples.
Each retrieval-remix cycle enabled users to explore alternatives UI designs and make more confident adjustments, \rev{reflecting a design process that participants perceived as more structured and effective.}
This iterative dynamic aligns with Schön's notion of reflection-in-action~\cite{sengers2005reflective, schon2017reflective}, where design understanding emerges through reciprocal feedback loops between thought and material form rather than from a single act of specification.

\sys{} supports this process by providing two complementary levels of exploration: global remix (whole-interface adaptation) and local remix (component-level refinement). 
\rev{This design allowed users to shift between divergent and convergent modes of thinking~\cite{guilford1950creativity}, broadly exploring stylistic directions before zooming in on specific elements to refine.}
This flexible structure mirrors how expert designers alternate between high-level concepts and detailed refinements~\cite{cross2004expertise}, \rev{helping non-designers navigate between abstraction levels in a way they described as manageable.}

These iterative processes also highlight the changing nature of the human–AI relationship during design.
More broadly, our findings point toward a reframing of AI's role in end user design support, i.e.,  from a generator that executes instructions to a reflective collaborator that scaffolds understanding.
\rev{Rather than assuming users can articulate their goals clearly upfront, \sys{} supported exploratory interactions through which participants reported clarifying what they wanted.}
In this view, AI's value lies not in automating design creation but in mediating reflection, i.e., providing traceable examples, supporting iteration, and prompting users to reason about their own choices.
This perspective invites a shift from optimizing generative accuracy toward designing systems that foster reflective efficacy --- how effectively users can examine and refine their ideas through collaboration with AI.

\rev{While \sys{} was experienced as empowering users' design process by fostering exploration, reflection, and confidence, these process-level benefits did not translate into significantly higher expert-rated output quality within the constrained study setting. This suggests that example-driven systems may primarily shape how users engage with design rather than immediately improving final outcomes, particularly in early-stage tasks. At the same time, reliance on retrieved examples raises potential risks of stylistic homogenization or over-dependence on existing designs if not carefully managed. Looking ahead, such systems may influence design practice by lowering entry barriers and shifting design toward a more exploratory, dialogue-based process, while also raising questions about originality, skill development, and long-term creative agency, which warrant further longitudinal investigation.}

\subsection{Design Takeaways}
Drawing on insights from our study, we outline several takeaways for designing example-driven AI systems for design that support end user creativity and trust.

\begin{itemize}
    \item \textbf{Scaffold articulation through example-grounded exploration}: Systems should treat examples not merely as outcomes to imitate but as articulation aids that help users articulate vague ideas. 
Interfaces that encourage users to search, browse, and iteratively remix examples can help them progressively clarify their design intent, even when their initial design goals are under-specified.
    \item \textbf{Make example provenance visible to foster informed trust}: Displaying source transparency cues, such as ratings, download counts, and developer information, can transform retrieved results into credible references rather than opaque suggestions. Such transparency allows users to calibrate trust and evaluate which examples best align with their needs.
    \item \textbf{Support reflective co-evolution between user and system}: Example-driven tools should emphasize iterative feedback loops that allow users to move fluidly between exploration and refinement. Providing complementary modes of interaction (e.g., global and local remix) helps users reason across abstraction levels, promoting structured yet creative iteration.
    \item \textbf{Reposition AI as a reflective collaborator, not a generator}: Future systems should value not only generative fidelity, i.e., how well the output matches user specifications, but also reflective efficacy, i.e.,  how effectively the system helps users interpret, evaluate, and refine their own ideas.  
AI can augment design thinking by mediating reflection: returning interpretable examples, revealing its retrieval rationale, and encouraging users to reframe their goals through exploration.
\end{itemize}

\subsection{Ethical Considerations}
Building example-driven AI systems for design requires careful attention to ethical and legal aspects of data sourcing, attribution, and user interaction.  
\sys{} relies on real-world UI examples retrieved from publicly available repositories to support design exploration.  
It is therefore essential to ensure that all referenced or displayed examples are collected and used in accordance with licensing agreements and data protection regulations.

Future systems that integrate similar retrieval mechanisms should prioritize datasets and sources that are explicitly open-licensed or permit fair-use display for research and educational purposes.  
When visual examples originate from commercial or user-generated platforms, clear attribution should be maintained to respect the intellectual property of original creators and promote responsible reuse.  
Displaying source metadata, as implemented in \sys{}, not only supports transparency and trust but also reinforces ethical visibility by acknowledging the provenance of each design.

Finally, as AI systems increasingly mediate creative processes, designers and researchers must be mindful of how example retrieval and remixing could influence originality and authorship.  
Ethical design assistance should aim to empower users' learning and reflection rather than encourage direct imitation.

\subsection{Limitations and Future Work}
Our study has several limitations. For the design tasks, we focused on creating a single mobile UI page based on a given context (e.g., a restaurant menu or a news app page).  
This focus allowed us to closely examine how end users engage in example-driven workflow within a bounded design space.  
While \sys{} is capable of generating interactive and multi-screen user interfaces, our evaluation was intentionally limited to static single-screen UIs due to time constraints and to maintain experimental control.  
Future work could extend \sys{} to support multi-screen or interactive design sequences, exploring how example-driven reflection extends across connected design contexts.

In addition, our evaluation was conducted as a single-session lab study focused on short-term design activities.  
While this setup allowed controlled observation of users' interactions with \sys{}, it does not capture how the system might be used in more naturalistic or collaborative settings.
Future evaluations could explore how users employ example-driven design support in real-world or longitudinal contexts, where goals and workflows are more diverse.

\rev{Besides, the scope of our evaluation opens up several opportunities for future investigation. Our study included participants with varied prior experience in AI-assisted UI design, but was not designed to explicitly contrast different end-user populations (e.g., users with versus without prior AI design experience). Future work could build on our findings by conducting broader user studies with more clearly defined target groups and by examining how prior experience shapes system use and perception. 
In addition, our study evaluated retrieval and remixing as an integrated workflow and did not isolate the individual contributions of these components. Future work could address this limitation through component-level ablation studies (e.g., retrieval-only, remix-only, and combined conditions) to more precisely characterize how each component supports exploration, reflection, and intent articulation.
Moreover, while our evaluation focused on short-term, task-based interactions within a single session, complementary evaluation approaches, such as longitudinal usage, learning trajectories, or changes in intent articulation over time, could provide deeper insight into how users engage with example-driven design tools in practice. Finally, source transparency cues were examined as part of the integrated system design; future studies could isolate and evaluate transparency features independently (e.g., with and without metadata cues) to more precisely characterize their role in supporting trust and design decision-making.}

Finally, while \sys{} effectively supports text and component-based retrieval, its current workflow is primarily user-initiated.  
Future work could explore integrating agentic mechanisms that enable the system to proactively suggest examples, refinements, or design directions based on users' ongoing actions and emerging goals.  
Such an extension would move \sys{} toward a more collaborative design partner --- capable of anticipating user needs and coordinating multi-step remix operations. 
\section{Conclusion}
In this paper, we introduced \sys{}, an interactive system that supports end users in designing mobile user interfaces through an example-driven design workflow. 
Grounded in multimodal retrieval-augmented generation, \sys{} enables users to iteratively search, select, and adapt real-world UI examples to refine their designs. 
By presenting source transparency cues such as ratings, download counts, and developer information of example UIs, the system helps users make more informed and trustworthy design choices. 
Our user study with 24 participants showed that \sys{} significantly improved users' ability to achieve their design goals, iterate effectively, and explore alternative UI designs.
Participants also reported that source transparency cues enhanced their confidence in adapting examples.
These findings highlight how transparent, example-driven interactions can advance a more reflective and trustworthy form of human–AI collaboration in end user design.
We hope the insights from this work inspire future research on developing example-driven AI systems that empower end users to design with greater creativity, confidence, and control.
\rev{Immediate directions for future work include extending example-driven support to multi-screen or interactive UI flows, exploring more proactive or agentic suggestion mechanisms, and examining longer-term use in real-world or collaborative settings.}

\section{GenAI Usage Disclosure}
In our study, AI assistants were used selectively and in accordance with the ACM Policy on the Use of Generative AI in Publications.  
We utilized ChatGPT and Grammarly for minor paraphrasing, grammar correction and code debugging.
These tools were applied minimally to ensure the authenticity of our work and to adhere strictly to ACM's ethical and publication standards.  
Our use of these AI tools was focused, responsible, and aimed at supplementing rather than replacing human input and expertise in our research process.
\begin{acks}
This project was made possible by ETH AI Center Doctoral Fellowships to Junling Wang, with partial support from the ETH Zurich Foundation. Additionally, the authors wish to thank the reviewers, members of the PEACH Lab at ETH Zurich, and the participants in the user study.
\end{acks}

\begin{figure*}[h]
    \centering
    \includegraphics[width=\textwidth]{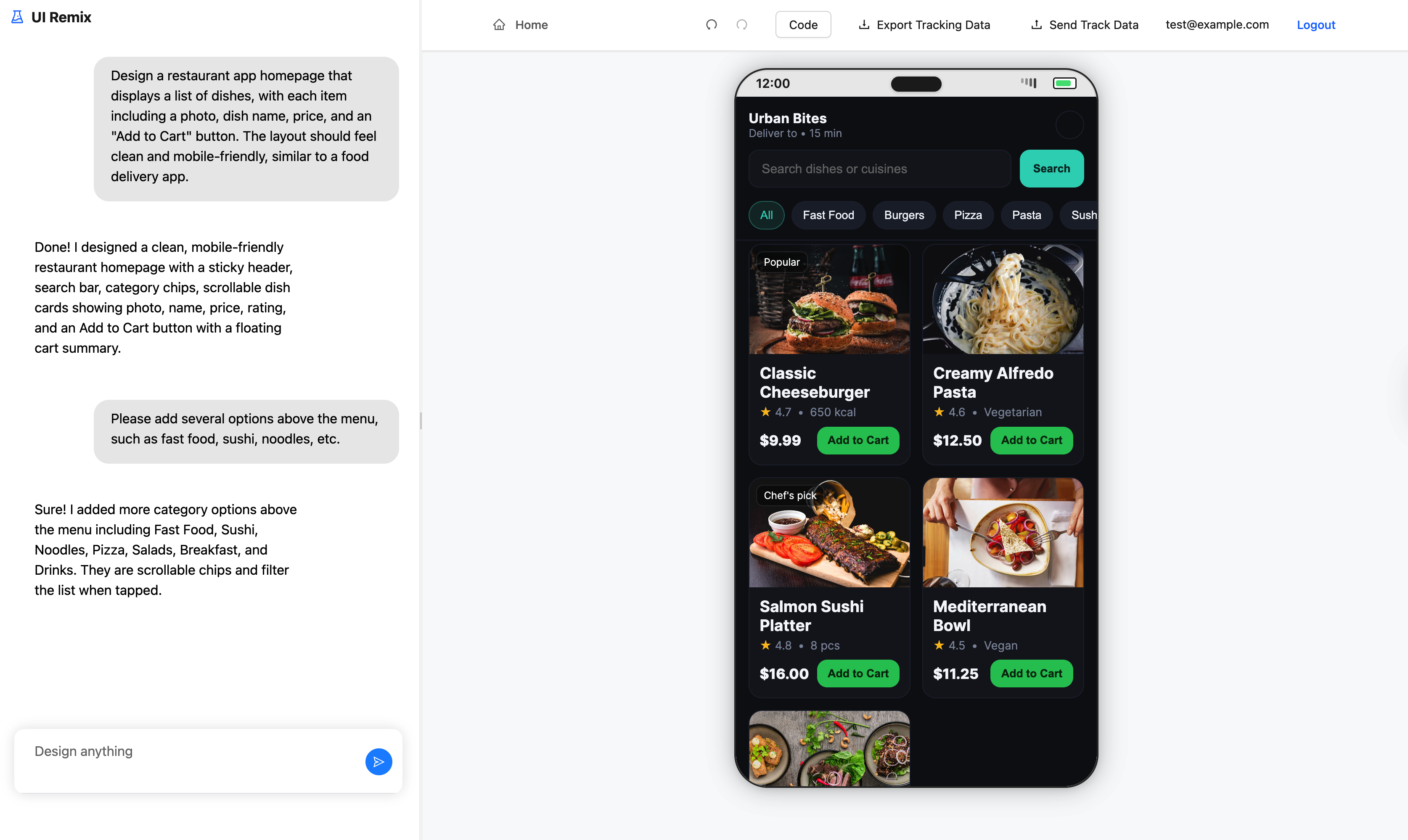}
    \vspace{-10pt}
    \caption{
    \rev{A screenshot of GPT-Canvas baseline system.}
}
    \label{fig:fig3}
\end{figure*}

\bibliographystyle{ACM-Reference-Format}
\bibliography{ref}
\appendix
\rev{\section{GPT-Canvas baseline Configurations}}
\rev{We present a screenshot of the GPT-Canvas baseline in Figure~\ref{fig:fig3}. The baseline adopts a design similar to GPT Canvas~\cite{openaiIntroducingCanvas} and Figma Make~\cite{figmaIntroducingFigma}: the left panel allows users to chat with the \ac{MLLM}, while the right panel mirrors the GPT Canvas layout, enabling users to preview the UI and switch back to the code view by clicking a button. The baseline uses the same \ac{MLLM} as \sys{}, but excludes the Example Gallery (Figure~\ref{fig:fig1}B) and the corresponding search and apply modes in the Conversation Panel (Figure~\ref{fig:fig1}A).}

\end{document}